\documentclass[12pt]{article}

\usepackage{latexsym}
\usepackage{amssymb,amsfonts,amsmath}
\usepackage{graphicx} 
\usepackage{indentfirst}
\usepackage{bbm}
\usepackage{amssymb}
\usepackage{verbatim}
\usepackage{amsmath, amsthm,amssymb}
\usepackage{mathrsfs}
\usepackage{hyperref}
\usepackage{amsfonts}
\usepackage{dsfont}
\usepackage{cite}
\usepackage{xcolor}
\usepackage[multiple]{footmisc}

\parskip=\medskipamount

\newcommand {\cD}{{\cal D}}
\newcommand {\cE}{{\cal E}}

\newcommand {\cL}{{\cal L}}
\newcommand {\cM}{{\cal M}}
\newcommand {\cN}{{\cal N}}

\newcommand {\cP}{{\cal P}}

\newcommand {\cV}{{\cal V}}
\newcommand {\cW}{{\cal W}}

\def\a{\alpha}

\def\b{\beta}

\def\d{\delta}
\def\e{\epsilon}

\def\g{\gamma}
\def\G{\Gamma}

\def\j{\psi}
\def\k{\kappa}
\def\l{\lambda}
\def\m{\mu}

\def\o{\omega}
\def\p{\pi}
\def\q{\theta}

\def\s{\sigma}
\def\t{\tau}

\def\z{\zeta}
\def\D{\Delta}
\def\F{\Phi}
\def\J{\Psi}
\def\L{\Lambda}
\def\O{\Omega}

\def\Q{\Theta}
\def\S{\Sigma}

\def\X{\Xi}

\def\rd{{\rm d}}
\def\ri{{\rm i}}
\def\re{{\rm e}}

\newcommand{\ad}{{\dot{\alpha}}}                           
\newcommand{\bd}{{\dot{\beta}}}                            
\newcommand{\ve}{\varepsilon}                            
\newcommand{\cDB}{{\bar\cD}}                            

\newcommand{\pa}{\partial}                           
\newcommand{\hf}{\frac12}

\newcommand{\vf}{\varphi}

\newcommand{\be}{\begin{equation}}
\newcommand{\ee}{\end{equation}}
\newcommand{\bea}{\begin{eqnarray}}
\newcommand{\eea}{\end{eqnarray}}
\newcommand{\non}{\nonumber}
\newcommand{\1}{{\underline{1}}}
\newcommand{\2}{{\underline{2}}}

\newcommand{\bm}[1]{\mbox{\boldmath$#1$}}

\def\double #1{#1{\hbox{\kern-2pt $#1$}}}

\newcommand{\gd}{{\dot\g}}

\newcommand{\qb}{{\bar{\theta}}}

\newif\ifdtup

\newcommand{\bsubeq}{\begin{subequations}}
\newcommand{\esubeq}{\end{subequations}}

\numberwithin{equation}{section}

\newcommand{\sSp}{\mathsf{Sp}}
\newcommand{\sSU}{\mathsf{SU}}
\newcommand{\sSL}{\mathsf{SL}}

\newcommand{\sU}{\mathsf{U}}

\begin{document}

\begin{titlepage}
\begin{flushright}
June, 2020 \\
\end{flushright}
\vspace{5mm}

\begin{center}
{\Large \bf 
Non-compact duality, super-Weyl invariance and effective actions}
\end{center}

\begin{center}

{\bf Sergei M. Kuzenko} \\
\vspace{5mm}

\footnotesize{
{\it Department of Physics M013, The University of Western Australia\\
35 Stirling Highway, Perth W.A. 6009, Australia}}  
~\\
\vspace{2mm}

\end{center}

\begin{abstract}
\baselineskip=14pt
In both ${\cal N}=1$ and ${\cal N}=2$ supersymmetry, it is known that $\mathsf{Sp}(2n, {\mathbb R})$ is the maximal duality group of $n$ vector multiplets coupled to chiral scalar multiplets $\tau (x,\theta) $ that parametrise the Hermitian symmetric space  $\mathsf{Sp}(2n, {\mathbb R})/ \mathsf{U}(n)$. If the coupling to  $\tau$ is introduced for $n$ superconformal gauge multiplets in a supergravity background, the action is also  invariant under super-Weyl transformations. Computing the path integral over the gauge prepotentials in curved superspace leads to an effective action $\Gamma [\tau, \bar \tau]$
with the following properties: (i)  its logarithmically divergent part is invariant under super-Weyl and rigid $\mathsf{Sp}(2n, {\mathbb R})$ transformations; (ii) the super-Weyl transformations are anomalous upon renormalisation. In this paper we describe the ${\cal N}=1$ and ${\cal N}=2$  locally supersymmetric ``induced actions'' which determine the  logarithmically divergent parts of the corresponding effective actions. In the ${\cal N}=1$ case, superfield heat kernel techniques are used to compute the induced action of a single  vector multiplet $(n=1)$ coupled to a chiral dilaton-axion multiplet. We also describe the general structure of ${\cal N}=1$ super-Weyl anomalies that contain weight-zero  chiral scalar multiplets $\Phi^I$ taking values in a K\"ahler manifold. Explicit anomaly calculations are carried out in the $n=1$ case.
\end{abstract}
\vspace{5mm}

\vfill

\vfill
\end{titlepage}

\newpage
\renewcommand{\thefootnote}{\arabic{footnote}}
\setcounter{footnote}{0}

\tableofcontents{}
\vspace{1cm}
\bigskip\hrule

\allowdisplaybreaks


\section{Introduction}

It is well known that  the group of electromagnetic duality rotations of free Maxwell's equations is  the compact group $\sU(1)$, assuming the duality invariance of the energy-momentum tensor. 
Almost forty years ago, it was shown by Gaillard and Zumino \cite{GZ1,Zumino} that 
the non-compact group $\sSp(2n, {\mathbb R})$ is the maximal duality group of 
$n$ vector field strengths $F_{ab}^I= -F_{ba}^I$ in the presence of scalar $\t^i$ parametrising the homogeneous 
space $\sSp(2n, {\mathbb R})/\sU(n)$.\footnote{The Gaillard-Zumino approach was inspired by patterns of duality in extended supergravity theories \cite{FSZ,CJ}, see
\cite{AFZ} for a review.}
In the absence of scalars, the largest duality group proves to be 
$\sU(n)$, the maximal compact subgroup of $\sSp(2n, {\mathbb R})$. 
These results admit a natural extension to the case when the pure vector field 
part  $L(F)$ of the Lagrangian $L(F,\t)$ is a nonlinear self-dual theory 
 \cite{GR1,GR2,GZ2,GZ3,AT} (see \cite{KT2,AFZ,Tanii} for reviews), for instance 
 the Born-Infeld theory. However, in the case that $L(F)$ is quadratic, the $F$-dependent part of $L(F,\t)$ is also invariant under the Weyl transformations in curved space. Then, computing the path integral over the gauge fields  leads 
 to an effective action, $\G [\t]$, such that its logarithmically divergent part 
 is invariant under Weyl and 
rigid $\sSp(2n, {\mathbb R})$ transformations, see, e.g., \cite{FT,RT} for formal arguments. Both symmetries are anomalous at the quantum level, but the logarithmically divergent part of the one-loop effective action is invariant under these transformations.
This simple observation offers
a powerful tool to construct higher-derivative actions with required symmetry properties as induced actions. For instance, 
this idea was employed  in an important paper 
by Buchbinder, Pletnev and Tseytlin \cite{BPT} to derive
the bosonic sector of $\cN=4$ conformal supergravity.\footnote{Some
of the relevant terms were missed in \cite{BPT}. The complete action for $\cN=4$ conformal supergravity was constructed in  \cite{CS,BCdeWS}.}

The analysis of Ref. \cite{BPT} was based in part on the earlier results of Osborn 
\cite{Osborn} who studied the effective action, $\G [\t, \bar \t]$,  
obtained by integrating out the quantum gauge field 
in the model with classical action
\bea 
L(F , \t, \bar \t) = - \frac 14 \re^{-\vf} F^{ab}F_{ab} +\frac 14 {\mathfrak a} \tilde F^{ab} F_{ab} ~, 
\qquad \t = {\mathfrak a} +\ri e^{-\vf}~.
\eea 
Here $\tilde F^{ab}$ is the Hodge dual of $F_{ab}$, 
and $\mathfrak a$ and $\vf$ are the axion and dilation, respectively.
The logarithmic divergence of the effective  was shown to have the form 
\bea
{\mathfrak L} &=&  \frac{1}{2( {\rm Im}\, \t)^2} \Big[ \cD^2 \t \cD^2 \bar \t 
- 2 (R^{ab}- \frac 13 \eta^{ab} R) \nabla_a \t \nabla_b \bar \t \Big] \non \\
&&+ \frac{1}{12 ( {\rm Im}\, \t)^4} 
\Big[ \a \nabla^a \t \nabla_a \t \nabla^b \bar \t  \nabla_b \bar \t 
+\b \nabla^a \t \nabla_a \bar  \t \nabla^b  \t  \nabla_b \bar \t 
\Big] 
\label{1.2}
\eea
where 
\bea
\cD^2 \t := \nabla^a \nabla_a \t + \frac{\ri}{{\rm Im}\, \t} \nabla^a \t \nabla_a \t ~,
\eea
and $\a$ and $\b$ are numerical parameters. 
The Lagrangian \eqref{1.2} is manifestly invariant under $\sSL(2,{\mathbb R}) $
transformations 
\bea
\t \to \t' = \frac{a\t + b}{c\t +d} ~, \qquad \left(
\begin{array}{cc}
 a  & b\\
c &   d 
\end{array}
\right) \in \sSL(2,{\mathbb R}) ~.
\label{1.4}
\eea
The functional $\int \rd^4 x \, e\, {\mathfrak L} $ 
proves to be Weyl invariant since
the scalar field $\t$ is inert under the Weyl transformations.
The Weyl invariance follows from the fact that the Fradkin-Tseytlin (FT)
operator \cite{FT1982}
\bea
\Delta_0 = (\nabla^a \nabla_a)^2 + 2 \nabla^a \big(
	 {R}_{ab} \,\nabla^b 
	- \tfrac{1}{3} {R} \,\nabla_a
	\big)
\label{1.5}
\eea
is conformal.\footnote{This operator was  re-discovered 
by Paneitz in 1983 \cite{Paneitz} and Riegert in 1984 \cite{Riegert}.}
 
Soon after Osborn's work \cite{Osborn} 
 several attempts were made  to extend his construction to supersymmetric case.
  So far no success has been achieved, 
 mainly due to the following two reasons. Firstly, the $\cN=1$ supersymmetric extension of the  FT
 operator was constructed only relatively recently
 \cite{ButterK13,BdeWKL} (and its $\cN=2$ cousin was presented in \cite{BdeWKL}).
 Secondly, the effective action corresponding to the vector multiplet model 
 \eqref{VM-action} proves to involve
 functional determinants of non-minimal 
differential operators, and these are  much harder to evaluate in superspace than 
in ordinary quantum field theory. One of the goals of this work is 
to provide such a supersymmetric generalisation of \cite{Osborn}.
Actually the scope of the present work is much broader, and in the remainder of this section we briefly describe the main results obtained below.

For every K\"ahler manifold $\cM$, we construct a  $\cN=1$ locally superconformal  
four-derivative action
which is formulated in terms of covariantly chiral superfields $\F^I$
 taking values in $\cM$. This action is invariant under (i) target-space isometries; and (ii) super-Weyl transformations of the supergravity multiplet. It  is given by  
 eq. \eqref{2.4} and involves two independent structures. The first term 
 in the right-hand side  of 
 \eqref{2.4} contains  a $\sigma$-model extension of 
 the $\cN=1$ supersymmetric  FT operator.
 The second term in \eqref{2.4} involves the Riemann curvature tensor of $\cM$.\footnote{$\cN=2$ supersymmetry uniquely fixes the
 relative coefficient $\a$  in \eqref{2.4} to be $\a =1$.} 
 In the case that $\cM$ is the Hermitian symmetric space $\sSp(2n, {\mathbb R})/\sU(n)$,
 \eqref{2.4} determines the  logarithmically divergent parts of the  effective action 
 for $n$ vector multiplets coupled to $\F$.\footnote{Several global parametrisations of  
 $\sSp(2n, {\mathbb R})/\sU(n)$ and the corresponding K\"ahler potential are described, e.g., in \cite{AFZ,Hua,AKL}.}

We also present 
the general structure of ${\cal N}=1$ super-Weyl anomalies that contain weight-zero  chiral multiplets $\F^I$
 taking values in $\cM$. The final expression for the anomaly is given by eq. 
 \eqref{N=1SWA5}. This anomaly is universal. Given a super-Weyl invariant theory 
 coupled to background sources $\F^I$, the anomaly  \eqref{N=1SWA5} 
 should characterise the effective action obtained by integrating out the quantum fields.
 Analogous anomaly results for the $\cN=0$ and $\cN=2$ cases are known in the literature  \cite{GHKSST,SchT2}, and this paper fills a gap concerning  $\cN=1$
 superconformal symmetry.

When dealing with $\cN=1$ local supersymmetry, 
me make use of the Grimm-Wess-Zumino geometry \cite{GWZ} which underlies the Wess-Zumino formulation \cite{WZ} for old minimal supergravity (see \cite{WB} for a review). Our two-component spinor notation and conventions follow \cite{BK}.
The algebra of supergravity covariant derivatives is given in the appendix. 
To describe $\cN=2$ (conformal) supergravity, 
the so-called $\sSU(2)$ superspace formulation \cite{KLRT-M} is used.


\section{Superconformal higher-derivative actions} 

In this section we describe $\cN=1$ and $\cN=2$ locally superconformal higher-derivative actions which are formulated in terms of chiral scalar superfields parametrising a K\"ahler manifold, 
for example the Hermitian symmetric spaces $\sSp(2n, {\mathbb R})/\sU(n)$.
Such actions occur, for instance,  as the logarithmically divergent part of effective actions in  
$\sSp(2n, {\mathbb R})$
duality-invariant theories.

\subsection{Local $\cN=1$ supersymmetry }

Let $\F$ be a covariantly chiral scalar  superfield, $\bar \cD^\ad \F =0$, which 
is neutral with respect to the super-Weyl transformations
\eqref{superweyl}, $\d_\s \F=0$.
It was demonstrated in \cite{BdeWKL} that the following functional
\bea
I &=& \frac{1}{16}  \int 
\rd^4x \rd^2 \q  \rd^2 \bar \q \, E \, 
\Big\{ \cD^2 \F \bar \cD^2 \bar \F -8 \cD^\a \F G_{\a\ad} \bar \cD^\ad \bar \F\Big\}
\eea
is super-Weyl invariant. The proof given in  \cite{BdeWKL} makes use of the fact that 
$I$ transforms into 
\bea
\d_\s I &=& \int 
\rd^4x \rd^2 \q  \rd^2 \bar \q \, E \, \Big\{ \cD^\a \s \cD_\a \F \bar \cD^2 \bar \F 
+4 \ri \cD_{\a\ad} \s \cD^\a \bar \cD^\ad \bar \F  + {\rm c.c.} \Big\} \non \\
&=& -\frac 18 \int 
\rd^4x \rd^2 \q  \rd^2 \bar \q \, E \, \bar \F \Big\{ \bar \cD^2 (\cD^\a \s \cD_\a \F )
+ 4\ri \bar \cD^\ad (\cD_{\a\ad} \s \cD^\a \F)\Big\} +{\rm c.c.}  =0~.
\eea
The higher-derivative superconformal action $I$ possesses non-trivial generalisations which will be described below.

Now we  assume, in addition, that 
$\F$ parametrises the upper half-plane, and therefore
$(\F -\bar \F)^{-1}$ exists.\footnote{The axion 
${\mathfrak a}(x) $ and dilaton 
$\vf(x)$ are the component fields of $\F(x,\q)$ defined by $\F|_{\q=0} = {\mathfrak a} +\ri \re^{-\vf}$.}
Consider the following locally supersymmetric action 
\bea
S &=& -\frac{1}{16}  \int 
\rd^4x \rd^2 \q  \rd^2 \bar \q \, 
E \, \frac{1}{(\F - \bar \F)^2} 
\Big\{ \nabla^2 \F \bar \nabla^2 \bar \F -8 \cD^\a \F G_{\a\ad} \bar \cD^\ad \bar \F\Big\}
\non \\
&& + \frac{\a}{8} \int 
\rd^4x \rd^2 \q  \rd^2 \bar \q \, 
E \, \frac{1}{(\F - \bar \F)^4} \cD^\a \F \cD_\a \F 
\bar \cD_\ad \bar \F \bar \cD^\ad \bar \F~,
\label{action2.3}
\eea
where $\a$ is a real parameter, and 
\bea
\nabla^2 \F = \cD^2 \F - 2 \frac{\cD^\a \F \cD_\a \F}{\F -\bar \F} ~, 
\qquad 
\bar \nabla^2 \bar \F = \bar \cD^2 \bar \F 
+ 2 \frac{\bar \cD_\ad \bar \F \bar \cD^\ad \bar  \F}{\F -\bar \F} 
\label{2.44}
\eea
are $\sSL (2, {\mathbb R}) $ covariant derivatives. The  action proves to be 
super-Weyl invariant. It is also invariant under 
 fractional linear $\sSL (2, {\mathbb R}) $ transformations 
\bea
\F \to \F' = \frac{a\F + b}{c\F +d} ~, \qquad \left(
\begin{array}{cc}
 a  & b\\
c &   d 
\end{array}
\right) \in \sSL(2,{\mathbb R}) ~.
\label{fractional_linear}
\eea

The above model can be generalised as follows. Let $K(\F^I, \bar \F^{\bar J})$ be the 
K\"ahler potential of a K\"ahler manifold $\cM$ .  We introduce a higher-derivative 
locally supersymmetric theory described in terms of covariantly chiral scalar superfields
$\F^I$, $\bar \cD^\ad \F^I=0$, which are neutral under the super-Weyl transformations, 
$\d_\s \F^I =0$. The  action is given by  
\bea
S &=& \frac{1}{16}  \int 
\rd^4x \rd^2 \q  \rd^2 \bar \q \, 
E \, g_{I \bar J} (\F, \bar \F) 
\Big\{ \nabla^2 \F^I \bar \nabla^2 \bar \F^{\bar J} 
-8 G_{\a\ad}\cD^\a \F^I  \bar \cD^\ad \bar \F^{\bar J} \Big\}\non \\
&&+ \frac{\a}{16} \int 
\rd^4x \rd^2 \q  \rd^2 \bar \q \, E\,R_{I \bar J K  \bar L} (\F, \bar \F)  \cD^\a \F^I \cD_\a \F^K \bar \cD_\ad 
\bar \F^{\bar J} \bar \cD^\ad \bar \F^{\bar L}
~,
\label{2.4}
\eea
where $g_{I\bar J} = \pa_I \pa_{\bar J} K$ is the K\"ahler metric, 
$R_{I \bar J K  \bar L} (\F, \bar \F)$ 
the Riemann curvature of the K\"ahler manifold, 
and 
\bea
\nabla^2 \F^I = \bar \cD^2 \F^I + \G^I_{KL} \cD^\a \F^K \cD_\a \F^L~.
\label{nabla2}
\eea
We recall that the Christoffel symbols $\G^I_{KL} $ and the curvature
$R_{I \bar J K  \bar L} $ are given by the expressions 
\bea
\G^I_{JK} = g^{I \bar L} \pa_J \pa_K \pa_{\bar L} K ~,
\quad R_{I \bar J K  \bar L} = \pa_I \pa_K \pa_{\bar J} \pa_{\bar L}  K 
-g^{M \bar N} \pa_I \pa_K \pa_{\bar N} K  \pa_{\bar J} \pa_{\bar L}  \pa_M K ~.
\eea
It may be shown that the action \eqref{2.4} is super-Weyl invariant. 
This action is manifestly invariant under K\"ahler transformations
\bea
K(\F, \bar \F) \to K(\F, \bar \F) + \L(\F) + \bar \L (\bar \F) ~,
\label{Kahler}
\eea
with $\L(\F) $ being an arbitrary holomorphic function.


\subsection{Local $\cN=2$ supersymmetry}

Let $ X$ be a covariantly chiral scalar superfield, $\bar \cD_i^\ad { X} =0$, defined to be invariant under the $\cN=2$ super-Weyl transformations \cite{KLRT-M}, $\d_\S { X}=0$, 
with the super-Weyl parameter $\S$ being  covariantly chiral, $\bar \cD_i^\ad {\S} =0$.
As in the $\cN=1$ case, we assume  that $({ X}-\bar { X})^{-1}$ exists. 
The following locally $\cN=2$ supersymmetric action 
\bea
S =  - \int 
\rd^4x \rd^4 \q  \rd^4 \bar \q \, 
E \, \ln \Big[\ri ( \bar{ X} - { X})\Big]
\label{2.5}
\eea
is obviously super-Weyl invariant. 
Moreover, it is invariant under 
$\sSL (2, {\mathbb R}) $ transformations 
\bea
 X \to { X}' = \frac{a{ X} + b}{c{ X} +d} ~, \qquad \left(
\begin{array}{cc}
 a  & b\\
c &   d 
\end{array}
\right) \in \sSL(2,{\mathbb R}) ~.
\eea
The superfield Lagrangian in the higher-derivative action \eqref{2.5} is proportional to the K\"ahler potential $K(X, \bar X ) = -  \ln \,{\rm i} (\bar X  -    X )$
of the Hermitian symmetric space $\sSL(2,{\mathbb R})/\sU(1)$.

The above model has a natural generalisation given by  
\bea
S =   \int 
\rd^4x \rd^4 \q  \rd^4 \bar \q \, 
E \, K(X^I, \bar X^{\bar J} )~, \qquad \bar \cD_i^\ad { X}^I =0~, \qquad 
\d_\s X^I=0~,
\label{2.8}
\eea
where the K\"ahler potential $K$ is the same as in \eqref{2.4}. The 
action is super-Weyl invariant if the chiral multiplets 
$X^I$ are  inert under the super-Weyl transformations. 
 Using the standard properties of $\sSU(2)$ superspace \cite{KLRT-M},
 it follows that the action \eqref{2.8} is super-Weyl invariant. 
Due to the important property \cite{Muller,KT-M09}
\bea
{\bar \cD}^{\ad}_i  \S=0
\quad \Longrightarrow \quad 
\int \rd^4 x \,{\rm d}^4\q\,{\rm d}^4{\bar \q}\,E\, \S=0~,
\eea
which holds for any covariantly chiral scalar $\S$, we observe that  the action \eqref{2.8} is also invariant under K\"ahler transformations
\bea
K(X, \bar X) \to K(X, \bar X) + \L(X) + \bar \L (\bar X) ~,
\eea
with $\L(X) $ being an arbitrary holomorphic function. It follows that \eqref{2.8} 
is invariant under  isometry transformations of the K\"ahler space.

Superconformal actions of the form \eqref{2.8} were discussed in \cite{BKT} in the rigid 
supersymmetric case,  and later in the supergravity 
framework \cite{deWKvZ,BdeWKL,GHKSST}. 
In Ref. \cite{BKT} $X$ was identified with the primary dimension-0 chiral scalar 
\bea
 X = W^{-2} \,{\bar D}^4 \ln {\bar W}~,
\eea
where $W$ is the field strength of a vector multiplet, $\bar D^\ad_i W =0$, 
and 
\bea
{\bar D}^4 = {1\over 16} 
({\bar D}_{\hat{1}} )^2 ({\bar D}_{\hat{2}})^2 
=  \frac{1}{48} {\bar D}^{ij}  {\bar D}_{ij}~, \qquad {\bar D}^{ij} 
= {\bar D}_\ad^{i} {\bar D}^{j \ad}~, \qquad i,j = \1, \2
\eea
is the chiral projection operator. The field strength $W$ is a reduced chiral superfield,
$D^{ij} W = \bar D^{ij} \bar{W}$.
For several Abelian field strengths $W^I$, higher-derivative actions
\bea
S =   \int 
\rd^4x \rd^4 \q  \rd^4 \bar \q \,  H(W^I, \bar W^{J} )
\eea
were considered, e.g.,  in \cite{Henningson,deWGR}. Locally supersymmetric 
extensions of such actions were analysed, e.g., in  \cite{deWKvZ}. 


\subsection{Relating the $\cN=2$ and $\cN=1$ actions}

The models \eqref{2.4} and \eqref{2.8} are intimately related to each other. 
Their relationship is most transparent in Minkowski superspace, with $D_\a^i$ and $\bar D^\ad_i$ being the corresponding spinor covariant derivatives, $i = \1, \2$.
Then the $\cN=2$ 
chiral superfield $X^I$, $\bar D^\ad_i X^I=0$, is equivalent to three $\cN=1$ chiral superfields $\F^I$, $\l^I_\a$ and $Z^I$  defined as follows:
\bea
\F^I :=X^I \big|_{\q_{\2}=0} ~, \qquad
\sqrt{2}  \O^I_\a := D^{\2}_\a X^I \big|_{\q_\2 =0}~, 
\qquad Z^I := -\frac 14 (D^\2)^2 X^I\big|_{\q_2=0} ~.
\eea
The $\cN=2$ supersymmetric action \eqref{2.8} reduces to $\cN=1$ Minkowski superspace
\bea
S =   \int  \rd^4x \rd^4 \q  \rd^4 \bar \q \, K(X, \bar X ) 
  = \frac{1}{16} \int \rd^4x \rd^2 \q  \rd^2 \bar \q \, 
  (D^\2)^2 (\bar D_\2)^2  K(X, \bar X )\Big|_{\q_\2 = \bar \q^\2= 0}~.
\eea
Now direct calculations give 
\bea
S &=& \frac{1}{16}  \int 
\rd^4x \rd^2 \q  \rd^2 \bar \q \,   \bigg\{ g_{I \bar J}
\nabla^2 \F^I \bar \nabla^2 \bar \F^{\bar J} 
+ R_{I \bar J K  \bar L}   D^\a \F^I D_\a \F^K \bar D_\ad 
\bar \F^{\bar I} \bar D^\ad \bar \F^{\bar L} \bigg\}  \non \\
&& + \int 
\rd^4x \rd^2 \q  \rd^2 \bar \q \,   g_{I \bar J} \bigg\{ {\mathbb Z}^I \bar {\mathbb Z}^{\bar J}
-{\ri} \O^{I \a} \nabla_{\a\ad} \bar \O^{\bar J \ad} \bigg\}~,
\label{2.14}
\eea
where we have defined
\bea
{\mathbb Z}^I = Z^I -\frac 14 \G^I_{JK} \O^{I\a} \O^K_\a ~, \qquad
\nabla_{\a\ad} \bar \O^{\bar I \ad} = \pa_{\a\ad} \bar \O^{\bar J \ad}
+\G^{\bar I}_{\bar J \bar K} \pa_{\a\ad}\bar \F^{\bar J} \bar \O^{\bar K \ad}~.
\eea
The superfield ${\mathbb Z}^I$ transforms as a target-space vector under holomorphic reparametrisations of the K\"ahler manifold,  but it is not chiral unlike $Z^I$.
The first term in \eqref{2.14} is the flat-superspace version of \eqref{2.4} with $a=b$.
Thus $\cN=2$ supersymmetry fixes the relative coefficient between the two structures
in \eqref{2.4}.

The curved superspace version of the quadratic spinor sector in \eqref{2.14} has recently been described in \cite{KPR}. That work introduced the model for a covariantly chiral spinor $\O_\a$ that is primary  of dimension 1/2, 
\bea
\bar \cD^\bd \O_\a =0~, \qquad \d_\s \O_\a = \hf \s \O_\a~.
\eea
The corresponding action 
\bea
S[\O, \bar \O] = -\int \rd^4x \rd^2 \q  \rd^2 \bar \q \, 
E \, \O^\a \Big( \ri \cD_{\a\ad} - G_{\a\ad} \Big) \bar \O^\ad 
\eea
is super-Weyl invariant, as follows from \eqref{superweyl} and \eqref{s-WeylG}.
This model is a special representative of the family of superconformal 
higher-derivative actions
\cite{KPR} formulated in terms of covariantly chiral rank-$n$ spinors 
$\O_{\a(n)}:= \O_{(\a_1 \dots \a_n)}$ of dimension $ (1-n/2)$, 
\bea
\bar \cD^\bd \O_{\a(n)} =0~, \qquad \d_\s \O_{\a(n)}  = \hf(2-n) \s \O_{\a(n)}~.
\eea


\section{Super-Weyl anomalies}

This section is devoted to ${\cal N}=1$ and $\cN=2$ super-Weyl anomalies that contain weight-zero superconformal chiral multiplets  parametrising a K\"ahler manifold $\cM$.


\subsection{$\cN=1$ super-Weyl anomalies} 

There are two different contributions to $\cN=1$ super-Weyl anomalies. 
One of them is given in terms of the 
supergravity multiplet. Modulo cohomologically trivial contributions, 
 the super-Weyl variation of the effective action is
\bea
\d_\s \G &=& 2(a-c) \int \rd^4 x \rd^2 \q \,\cE \,\s W^{\a\b\g} W_{\a\b\g} +{\rm c.c.}
\non \\
&&+2a  \int\rd^4 x \rd^2 \q \rd^2 \bar \q  \,E \, (\s+\bar \s)(G^aG_a +2R \bar R)~.
\label{N=1SWA1} 
\eea
The functional structures which contribute to the anomaly \eqref{N=1SWA1}
were contained in the chiral super-$b_4$ coefficient computed 
in 1983 by McArthur \cite{McA}.  His work was followed by the cohomological analysis 
of Bonora, Pasti and Tonin \cite{BonoraPT} who arrived at the same structures. 
The explicit calculations of the super-Weyl anomalies for the scalar and vector multiplets 
were carried out in \cite{BK86}.
Various effective actions generating the anomaly \eqref{N=1SWA1} were proposed
in \cite{BK-PLB88,SchT,ButterK13,KST}.

The second type of  super-Weyl anomalies 
is determined by local couplings in a superconformal field theory.
 In general, 
 such an 
 anomaly 
  is given by 
 \bea
 \d_\s \widetilde  \G &=& \frac{1}{16} \int 
\rd^4x \rd^2 \q  \rd^2 \bar \q \, 
E \, \s \bigg\{  K_{\bar I} (\F, \bar \F) 
\Big( \cD^2 \bar \cD^2 \bar \F^{\bar I} 
+8 \cD^\a (G_{\a\ad} \bar \cD^\ad \bar \F^{\bar I}) \Big) \non \\
&&-  8
  K_{\bar I \bar J } (\F, \bar \F) 
  \Big(\cD^{\a\ad} \bar \F^{\bar I} \cD_{\a\ad}\bar \F^{\bar J} 
-\bar R \bar \cD_\ad \bar \F^{\bar I} \cD^\ad \bar \F^{\bar J} \Big) \bigg\} +{\rm c.c.} \non \\
&& + \frac{\a}{16} \int 
\rd^4x \rd^2 \q  \rd^2 \bar \q \, 
E \, (\s+\bar \s)
R_{I \bar J K  \bar L} (\F, \bar \F)  \cD^\a \F^I \cD_\a \F^K \bar \cD_\ad 
\bar \F^{\bar J} \bar \cD^\ad \bar \F^{\bar L}~,
\label{N=1SWA2} 
\eea
compare with  the $\cN=2$ super-Weyl anomaly \eqref{2.18}.
This anomaly obeys the Wess-Zumino consistency condition,
$[\d_{\s_2} , \d_{\s_1} ]\widetilde\G =0$,
 since 
\bea
\d_{\s_2} \d_{\s_1} \widetilde\G =0~,
\eea 
as a consequence of the following identity \cite{ButterK13}:
\bea
\d_\s \Big\{ \cD^2 \bar \cD^2 \bar \F + 8 \cD^\a (G_{\a\ad}\bar \cD^\ad \bar \F)\Big\}
&=& (\s +\bar \s)  \Big\{ \cD^2 \bar \cD^2 \bar \F 
+ 8 \cD^\a (G_{\a\ad} \bar \cD^\ad\bar \F)\Big\}
\non \\
&& 2 \bar \cD_\ad \Big\{ (\bar \cD^\ad \bar \F) \cD^2 \s 
+ 4\ri (\cD^{\a\ad} \bar \F) \cD_\a \s \Big\} ~.
\label{2.21}
\eea

At this point an important comment should be made.
It follows from \eqref{2.21} that the operator 
\bea
\D_{\rm c}\bar \F := -\frac{1}{64} (\bar \cD^2 -4R ) \Big\{ \cD^2 \bar \cD^2 \bar \F 
+ 8 \cD^\a (G_{\a\ad}\bar \cD^\ad \bar \F)\Big\}~, \qquad 
\bar \cD^\ad \D_{\rm c}\bar \F =0
\label{Delta}
\eea
is superconformal in the sense that it has the super-Weyl transformation law
\bea
\d_\s \D_{\rm c}\bar \F = 3\s \D_{\rm c}\bar \F ~.
\eea
$\D_{\rm c}$ is the $\cN=1$ supersymmetric extension of the Fradkin-Tseytlin operator \eqref{1.5}. This operator played important roles in the analyses 
carried out in \cite{ButterK13,LOR1}.

The super-Weyl anomaly \eqref{N=1SWA2} may be rewritten in a different form, 
 \bea
 \d_\s \widetilde\G &=& \frac{1}{16} \int 
\rd^4x \rd^2 \q  \rd^2 \bar \q \, 
E \,  \bigg\{ \s K_{\bar I} (\F, \bar \F) 
\Big( \cD^2 \bar \cD^2 \bar \F^{\bar I} 
+8 \cD^\a (G_{\a\ad} \bar \cD^\ad \bar \F^{\bar I}) \Big) \non \\
&&\qquad +   \cD^2 \Big( \s K_{\bar I \bar J } (\F, \bar \F) \Big) 
\bar  \cD_{\ad} \bar \F^{\bar I} \bar \cD^{\ad}\bar \F^{\bar J} 
 \bigg\} 
+{\rm c.c.} \non \\
&& + \frac{\a}{16} \int 
\rd^4x \rd^2 \q  \rd^2 \bar \q \, 
E \, (\s+\bar \s)
R_{I \bar J K  \bar L} (\F, \bar \F)  \cD^\a \F^I \cD_\a \F^K \bar \cD_\ad 
\bar \F^{\bar I} \bar \cD^\ad \bar \F^{\bar L}
~,
\label{N=1SWA3} 
\eea
which is more useful for analysing the behaviour of $ \d_\s \widetilde \G$ 
under  K\"ahler transformations \eqref{Kahler}. One obtains
\bea
\d_\L  \d_\s  \widetilde\G &=& \frac{1}{16} \int 
\rd^4x \rd^2 \q  \rd^2 \bar \q \, 
E \,  \bigg\{ \s\bar  \L_{\bar I} (\bar \F) 
\Big( \cD^2 \bar \cD^2 \bar \F^{\bar I} 
+8 \cD^\a (G_{\a\ad} \bar \cD^\ad \bar \F^{\bar I}) \Big) \non \\
&&\qquad +  ( \cD^2 \s )\bar \L_{\bar I \bar J } (\bar \F) 
\bar  \cD_{\ad} \bar \F^{\bar I} \bar \cD^{\ad}\bar \F^{\bar J} 
 \bigg\} 
+{\rm c.c.} \non \\
&=& \frac{1}{16} \int 
\rd^4x \rd^2 \q  \rd^2 \bar \q \, E \, \bar \L(\bar \F) \Big( 
\bar \cD^2 \cD^2 \s - 8 \bar \cD^\ad ( G_{\a\ad} \cD^\a \s) \Big)  +{\rm c.c.}
\label{KahlerWeyl}
\eea
Making use of the identity \cite{ButterK13}
\bea
\d_\s \Big\{ G^aG_a &+& 2R \bar R  - \frac{1}{4} \bar \cD^2 \bar R    \Big\}  = 
(\s+\bar \s)  \Big\{ G^aG_a + 2R \bar R  - \frac{1}{4} \bar \cD^2 \bar R  \Big\}  \non \\
&-&\frac{1}{16}\Big(\bar \cD^2 \cD^2  \s
-8 \bar \cD^\ad (G_{\a\ad}    \cD^\a  \s) \Big) 
+\hf  \cD^\a \Big(  R  \cD_\a  \s
- G_{\a\ad}  \bar  \cD^\ad \bar \s\Big)~,
\eea
it follows that  \eqref{KahlerWeyl} can be recast as the super-Weyl variation 
of a local functional, 
\bea
\d_\L  \d_\s \widetilde\G &=&- \d_\s \int 
\rd^4x \rd^2 \q  \rd^2 \bar \q \, E \,  \L( \F) 
 \Big( G^aG_a + 2R \bar R  - \frac{1}{4}  \cD^2 R    \Big) +{\rm c.c.} \non \\
 &=&- \d_\s \d_\L \int 
\rd^4x \rd^2 \q  \rd^2 \bar \q \, E \,  K( \F , \bar \F ) 
 \Big\{ G^aG_a + 2R \bar R  - \frac{1}{4} (  \cD^2 R +\bar \cD^2 \bar R )  \Big\} ~,
 \eea
 and therefore 
 \bea
 \d_\L \d_\s \bigg\{ \widetilde \G 
+ \int 
\rd^4x \rd^2 \q  \rd^2 \bar \q \, E \,  K( \F , \bar \F ) 
 \Big( G^aG_a + 2R \bar R  - \frac{1}{4} (  \cD^2 R +\bar \cD^2 \bar R )  \Big) 
 \bigg\}=0~.
 \eea

If we now introduce the functional 
\bea
\widetilde{\widetilde \G} := \widetilde \G
+ \int 
\rd^4x \rd^2 \q  \rd^2 \bar \q \, E \,  K( \F , \bar \F ) 
 \Big( G^aG_a + 2R \bar R  - \frac{1}{4} (  \cD^2 R +\bar \cD^2 \bar R )  \Big) ~,
 \eea
 then its super-Weyl variation may be shown to be 
 \bea
 \d_\s \widetilde{\widetilde \G} &=&-  \int 
\rd^4x \rd^2 \q  \rd^2 \bar \q \, E \, \s 
g_{I \bar J} (\F, \bar \F) 
\bigg\{ \frac{1}{16} \nabla^2 \F^I \bar \nabla^2 \bar \F^{\bar J} 
+ G_{\a\ad}\cD^\a \F^I  \bar \cD^\ad \bar \F^{\bar J} \non \\
&& -\frac{\ri}{2} \cD^\a \F^I \nabla_{\a\ad} \bar \cD^\ad \bar \F^{\bar J}  \bigg\}
+{\rm c.c.} \non \\
&&+ \frac{\a-1}{16} \int 
\rd^4x \rd^2 \q  \rd^2 \bar \q \, E \, (\s+\bar \s)
R_{I \bar J K  \bar L} (\F, \bar \F)  \cD^\a \F^I \cD_\a \F^K \bar \cD_\ad 
\bar \F^{\bar J} \bar \cD^\ad \bar \F^{\bar L}~,
\label{N=1SWA4} 
 \eea
where $\nabla^2 \F^I$ is given by  \eqref{nabla2}, 
and $\nabla_{\a\ad} \bar \cD^\ad \bar \F^{\bar I} $ is defined as 
\bea
\nabla_{\a\ad} \bar \cD^{\ad} \bar \F^{\bar I}
= \cD_{\a\ad} \bar \cD^{\ad} \bar \F^{\bar I}
+\G^{\bar I}_{\bar J \bar K} \cD_{\a\ad}\bar \F^{\bar J}  \bar \cD^{\ad} \bar \F^{\bar K}~.
\eea
The super-Weyl anomaly \eqref{N=1SWA4} is manifestly K\"ahler invariant.

Finally, we can recast the anomaly \eqref{N=1SWA4} in a more compact form by making use of the identity 
\bea
\d_\s  \int \rd^4x \rd^2 \q  \rd^2 \bar \q \, E \, 
g_{I \bar J}   G_{\a\ad}\cD^\a \F^I  \bar \cD^\ad \bar \F^{\bar J} 
&=& -\ri \int 
\rd^4x \rd^2 \q  \rd^2 \bar \q \, E \, \s 
g_{I \bar J} \cD^\a \F^I \nabla_{\a\ad} \bar \cD^\ad \bar \F^{\bar J}  \non \\
 + \frac 14  \int 
\rd^4x \rd^2 \q  \rd^2 \bar \q \, E \, \s 
\bigg\{ g_{I \bar J}  \nabla^2 \F^I \bar \nabla^2 \bar \F^{\bar J} 
&+&R_{I \bar J K  \bar L}  \cD^\a \F^I \cD_\a \F^K \bar \cD_\ad 
\bar \F^{\bar J} \bar \cD^\ad \bar \F^{\bar L} \non \\
&+&4 g_{I \bar J}   G_{\a\ad}\cD^\a \F^I  \bar \cD^\ad \bar \F^{\bar J} \bigg\} + {\rm c.c.}
\eea
This identity shows that the functional 
\bea
\G:=  \widetilde{\widetilde \G} + \hf   \int \rd^4x \rd^2 \q  \rd^2 \bar \q \, E \, 
g_{I \bar J} (\F , \bar \F)   G_{\a\ad}\cD^\a \F^I  \bar \cD^\ad \bar \F^{\bar J} 
 \eea
has the following super-Weyl variation 
\bea
 \d_\s \G &=&\frac{1}{16}\int  \rd^4x \rd^2 \q  \rd^2 \bar \q \, E \, \s 
g_{I \bar J} (\F, \bar \F) 
\Big\{  \nabla^2 \F^I \bar \nabla^2 \bar \F^{\bar J} 
-8  G_{\a\ad}\cD^\a \F^I  \bar \cD^\ad \bar \F^{\bar J} \Big\}
+{\rm c.c.} \non \\
&&+ \frac{\a+1}{16} \int 
\rd^4x \rd^2 \q  \rd^2 \bar \q \, E \, (\s+\bar \s)
R_{I \bar J K  \bar L} (\F, \bar \F)  \cD^\a \F^I \cD_\a \F^K \bar \cD_\ad 
\bar \F^{\bar J} \bar \cD^\ad \bar \F^{\bar L}~.~~~
\label{N=1SWA5} 
\eea
This is our final expression for the super-Weyl anomaly
determined by local couplings in $\cN=1$ superconformal field theories.


\subsection{$\cN=2$ super-Weyl anomalies} 

It is a curious fact that, in the   literature, 
the general structure of $\cN=2$ super-Weyl anomalies
has been understood better than in the $\cN=1$ case. 
Unlike the $\cN=1$ case, however,  no supergraph calculation of 
the $\cN=2$ super-Weyl anomalies has yet appeared. 

There are two different contributions to the $\cN=2$ super-Weyl anomaly. 
One of them is given in terms of the 
supergravity multiplet \cite{K13} and has the form 
\bea
\d_\S \G = (c-a) \int \rd^4x\, \rd^4\q\, \cE\, \S W^{\a\b}W_{\a\b} 
+ a \int \rd^4x\, \rd^4\q\, \cE\, \S \X ~+~{\rm c.c.} ~,
\label{N=2SWA1} 
\eea
for some  anomaly coefficients $a$ and $c$. 
Within the $\sSU(2)$ superspace approach \cite{KLRT-M},
the super-Weyl parameter $\S$ is an unrestricted chiral superfield.
The  chiral rank-2 spinor
$W_{\a\b} =W_{\b\a} $ in \eqref{N=2SWA1} 
 is the $\cN=2$ super-Weyl tensor. 
The composite chiral scalar $\X$ in the second term of \eqref{N=2SWA1} 
is the $\cN=2$ counterpart of the local operator \cite{ButterK13}
\bea 
\hat Q =  -\frac{1}{4} (\bar \cD^2 -4R ) \Big\{ G^aG_a + 2R \bar R  
- \frac{1}{4} \cD^2 R \Big\}
\eea
in $\cN=1$ supergravity.\footnote{$\hat Q$ 
is known in the literature as the $\cN=1$ supersymmetric $Q$-curvature \cite{LOR1,NN}.}
 The composite $\X$ has several fundamental  properties \cite{BdeWKL,K13}
which are also important  for the construction 
of $ \mathcal{N}  = 2$ Liouville SCFT in four dimensions \cite{LOR2}.

The other sector of the $\cN=2$ super-Weyl anomaly
is determined by local couplings in a superconformal field theory.  
According to \cite{GHKSST,SchT2}, it is given by 
\bea
\d_\S \G = \int \rd^4 x \,{\rm d}^4\q\,{\rm d}^4{\bar \q}\,E\, \big(\S +\bar \S \big)
K(X, \bar X) ~,
\label{2.18}
\eea
where the K\"ahler potential $K(X, \bar X) $ is the same as in \eqref{2.8}.
Since the chiral scalars $X^I$ are inert under the super-Weyl transformations, the anomaly clearly satisfies the Wess-Zumino consistency condition. 
The right-hand side of \eqref{2.18} is not invariant under K\"ahler transformations, 
unlike the $\cN=1$ super-Weyl anomaly \eqref{N=1SWA5}.
However, the $\cN=2$ super-Weyl anomaly is invariant under a joint K\"ahler-Weyl transformation. 
 A detailed analysis 
of the anomaly \eqref{2.18} is given in the original publications
 \cite{GHKSST,SchT2} to which the reader is referred for the technical details.
In the case of the Hermitian symmetric spaces  $\mathsf{Sp}(2n, {\mathbb R})/ \mathsf{U}(n)$, the K\"ahler potential $K(X, \bar X) $ can be chosen to be a globally defined function on the target space (see, e.g., \cite{AKL}). Choosing a different globally defined K\"ahler potential corresponds to a different scheme.


\section{Quantisation} \label{section4}

We consider the model for a massless vector multiplet coupled to a dilaton-axion chiral superfield $\F$, $\bar \cD^\ad \F=0$, 
in curved superspace. 
Its dynamics is described by the action  
\bea
S [V; \F , \bar \F]= - \frac{\ri }{4} \int \rd^4x \rd^2 \q  \, \cE \, \F W^\a (V)W_\a(V) +{\rm c.c.}~, 
\label{VM-action}
\eea
where $ W_\a(V) = -\frac 14 (\bar \cD^2 -4R) \cD_\a V$.
The chiral field strength $W_\a(V)$ and the action are invariant under gauge transformations
\bea
\d_\l V =\l + \bar \l ~, \qquad \bar \cD^\ad \l =0~.
\eea
The gauge prepotential $V $ is chosen to be super-Weyl inert, $\d_\s V=0$, which 
implies 
\bea
\d_\s W_\a (V) =\frac 32 \s W_\a(V)~.
\eea
As a consequence, the action \eqref{VM-action} is super-Weyl invariant. 
In addition, the model possesses $\sSL(2,{\mathbb R})$ duality.\footnote{Within the
$\cN=1$ Poincar\'e supersymmetry,  $\sSL(2,{\mathbb R})$ duality invariant couplings 
of the dilaton-axion multiplet to general models for self-dual  supersymmetric nonlinear electrodynamics were described in \cite{KT2}, while the case of the supersymmetric 
Born-Infeld action \cite{CF,BG} was first considered in \cite{BMZ}. The results of \cite{KT2} were generalised to supergravity  in  \cite{KMcC}.} 
The duality group acts on $\F$ by fractional linear transformations 
\eqref{fractional_linear}.

Making use of the chiral action rule 
\bea
\int\rd^4x\rd^2\q\rd^2\qb\, E \,\cL
= -\frac 14 \int\rd^4x\rd^2\q\, \cE \,\big(\cDB^2-4R\big)
\cL~,
\eea
the action \eqref{VM-action}  can be rewritten in the form 
\bea
S &=& \frac{1}{16} \int\rd^4x\rd^2\q\rd^2\qb\, E \, V \Big\{ 
\X \cD^\a (\bar \cD^2-4R) \cD_\a 
\non  \\
&& 
+ (\cD^\a \X) (\bar \cD^2 -4R) \cD_\a 
+ (\bar \cD_\ad \X) ( \cD^2 -4 \bar R) \bar \cD^\ad \Big\}V~,
\qquad \X \equiv \ri (\bar \F - \F)~.
\eea
We choose the following gauge-fixing term
\bea
S_{\rm G.F.} &=& - \frac{1}{16} \int\rd^4x\rd^2\q\rd^2\qb\, E \, \X
\big[(\bar \cD^2-4R) V\big]  \big[(\cD^2-4\bar R)V\big]  ~.
\eea
Then the  complete action is given by 
\bea
S + S_{\rm G.F.} &=&-  \hf \int\rd^4x\rd^2\q\rd^2\qb\, E \, 
V {\bm \D}_{\rm v} V~,
\eea
where we have introduced the second-order operator
\bea
{\bm \D}_{\rm v} = \X {\bm \Box}_{\rm v} 
&-&\frac{\ri}{4} (\bar \cD^\ad \X) \big\{\cD_{\a\ad} , \cD^\a \big\} 
-\frac{\ri}{4} ( \cD^\a \X) \big\{\cD_{\a\ad} , \bar \cD^\ad \big\} 
\non \\
&+&\frac{1}{16} (\bar \cD^2 \X) (\cD^2 -4 \bar R) 
+\frac{1}{16} ( \cD^2 \X) (\bar \cD^2 -4  R) \label{3.9} \\
&-& \hf (\cD^\a \X) (\cD_\a R) - \hf (\bar \cD_\ad \X) (\bar \cD^\ad \bar R) ~,
\non
\eea
in which  the operator $\Box_{\rm v}$ in the first term is
\bea
\Box_{\rm v} &=& -\frac 18  \cD^\a (\bar \cD^2-4R) \cD_\a 
+\frac{1}{16}\big\{ \bar \cD^2-4R ,  \cD^2-4\bar R \big\}  \non \\
&=& \cD^a \cD_a -\frac 14 G^{\a\ad} \big[ \cD_\a , \bar \cD_\ad \big]
-\frac 14 (\cD^\a R) \cD_\a - \frac 14 (\bar \cD_\ad \bar R) \bar \cD^\ad 
\label{operator4.9} \\
&&-\frac 14 (\cD^2R) - \frac 14 (\bar \cD^2 \bar R) + 2 R \bar R~. \non
\eea
The operator ${\bm \D}_{\rm v} $ reduces to $\Box_{\rm v}$ in the   $\X =1$ case.

A useful gauge condition is
\bea
\k(V) = -\frac 14  (\bar \cD^2-4R) V +\eta  ~, \qquad \bar \cD^\ad \eta=0
\eea
with $\eta$ a background chiral superfield. Then the Faddeev-Popov 
operator is $H^{(R)}$, where, in general,  $H^{(\j)} $ denotes the following operator
\bea
H^{(\j)} =  \left(
\begin{array}{cc}
 \j   &  - \frac 14 (\bar \cD^2 -4R) \ \\
 - \frac 14 (\cD^2 -4\bar R)  &   \bar \j
\end{array}
\right)~,
 \qquad \bar \cD^\ad \j =0 ~. 
 \eea
For the effective action we obtain
\begin{align}
{\rm e}^{\ri \G_{\rm v} [\F, \bar \F]} &=\int {\mathsf D} V 
\d_{+} \Big[ \eta - \frac 14  (\bar \cD^2-4R) V \Big] 
\d_{-}  \Big[ \bar \eta - \frac 14  ( \cD^2-4\bar R) 
V \Big]\, {\rm Det} \,H^{(R)} 
\re^{\ri S[V; \F, \bar \F] }~,
 \end{align}
 where ${\rm Det} \,H^{(R)} $ stands for the ghost determinant.
 Since the right-hand side is independent of $\eta$ and $\bar \eta$, we can average it 
 over these superfields with weight 
 \bea
{ \rm Det}^{1/2} {\bm H}_\X \, \re^{-\ri S [\eta , \bar \eta ; \X]}~, \qquad
S [\eta , \bar \eta ; \X]=  \int 
\rd^4x \rd^2 \q  \rd^2 \bar \q \, 
E \, \X \,\bar \eta  \eta ~,
\eea
where ${\bm H}_\X $ denotes the following operator
\bea
{\bm H}_\X=\left(
\begin{array}{cc}
 0   &  - \frac 14 (\bar \cD^2 -4R)\, \X \\
 - \frac 14 (\cD^2 -4\bar R)\, \X &   0 
\end{array}
\right) ~.
 \label{H_X} 
 \eea
 The operator ${\bm H}_\X$ is defined to act on the space  of column-vectors 
 \bea
 \left(
\begin{array}{c}
\eta  \\
 \bar \eta
\end{array}
\right) ~, \qquad \bar \cD^\ad \eta =0 
\non 
\eea
such that 
\bea
{\bm H}_\X  \left(
\begin{array}{c}
\eta  \\
 \bar \eta
\end{array}
\right) 
= \left(
\begin{array}{c}
   - \frac 14 (\bar \cD^2 -4R) (\X \bar \eta) \\
 - \frac 14 (\cD^2 -4\bar R)(\X \eta)  
\end{array}
\right) 
\equiv \left(
\begin{array}{c}
\cP^{\X}_{+-} \bar \eta  \\
 \cP^\X_{-+}  \eta
\end{array}
\right) ~.
\label{H+-}
\eea

The quantisation procedure described leads to the following representation for the effective action 
\bea
{\rm e}^{\ri \G_{\rm v} [\F, \bar \F]} &=&
 {\rm Det} \,H^{(R)} \,{\rm Det}^{1/2} {\bm H}_\X \, 
\int {\mathsf D} V 
\exp \bigg\{ -  \frac{\ri}{2} \int\rd^4x\rd^2\q\rd^2\qb\, E \, 
V {\bm \D}_{\rm v} V\bigg\}~,
\eea
and therefore
\begin{subequations}
\bea
\G_{\rm v} [\F, \bar \F] &=& \frac{\ri}{2} {\rm Tr} \, \ln {\bm \D}_{\rm v}\,
-\ri  {\rm Tr} \, \ln H^{(R)}  -\frac{\ri}{2}  {\rm Tr} \, \ln {\bm H}_\X 
\eea
or equivalently 
\bea
\G_{\rm v} [\F, \bar \F] &=& \G_{\rm v} +  \frac{\ri}{2} {\rm Tr} \, \ln \frac{{\bm \D}_{\rm v}}{\Box_{\rm v} }
-\frac{\ri}{2}  {\rm Tr} \, \ln \frac{{\bm H}_\X}{H^{(0)}  }~.
\label{3.16b}
\eea 
\end{subequations}
Here $ \G_{\rm v} $ is the effective action for the vector multiplet in a supergravity background \cite{BK86,BK}. It is obtained from $\G_{\rm v} [\F, \bar \F]$ by setting 
$\F=\ri$. The functional $ \G_{\rm v} $ was studied 
in \cite{BK86,BK}. It is the second and third terms in \eqref{3.16b} which contain the dependence of $\G_{\rm v} [\F, \bar \F] $ on $\F$ and $\bar \F$. The latter functionals will be studied in the next sections. 


\section{Heat kernel calculations (I)} \label{section5}

Heat kernel techniques in curved superspace have been developed by many authors 
over several decades  
\cite{McA,GNSZ,BK86,BK88,BKS,Butter2009,Leung1,Leung2}, see \cite{BK} for a review. What is special about the contributions in the second and third terms of \eqref{3.16b}, is that they involve non-minimal second-order differential operators 
for which 
the standard superfield Schwinger-DeWitt technique \cite{BK86,BK} 
is not directly applicable.

\subsection{Generalised Schwinger-DeWitt representation} 

 Associated with a second-order differential operator of the form 
 \bea
 { \D} = \O^{AB} \cD_B \cD_A + \Q^A\cD_A + \k= \hat { \D} + \k~, 
 \qquad \O^{AB} = (-1)^{\e_A \e_B} \O^{BA}
 \eea
 is the heat kernel $U (z,z'|s) $ which satisfies the equation 
\begin{subequations}\label{HeatKernel}
 \bea
 \Big( \ri \frac{\pa}{\pa s} + { \D} \Big) U(z,z'|s) =0
 \label{HeatKernel_a}
 \eea
 and the initial condition 
 \bea
 \lim_{s\to 0} U(z,z'|s) = E^{-1} \d^4(x-x') \d^2(\q-\q') \d^2 (\bar \q - \bar \q') 
 \equiv \d^{4|4} (z,z')~.
 \label{InitialCondition1}
 \eea
 \end{subequations}
 Keeping in mind the structure of the operator ${\bm \D}_{\rm v}$, eq. \eqref{3.9},
 it will be assumed that 
 \bea
 \O^{ab} = \O(z) \eta^{ab}~,
 \eea
 with $\eta_{ab} $ the Minkowski metric and $\O$ a nowhere vanishing background real superfield.
 
 We look for a solution to \eqref{HeatKernel} by making the ansatz 
 \bea
 U(z,z'|s) = - \frac{\ri}{ (4\p s)^2} \re^{\ri \s(z,z') / 2 s} 
 \sum_{n=0}^{\infty} a_n(z,z') (\ri s)^n ~,
 \label{5.44}
 \eea
 where the symmetric bi-scalars $\s$ and $a_n$ obey the equations
 \begin{subequations} \label{recurrence} 
 \bea
 && \O^{AB} \cD_B \s \cD_A \s = 2\s~,\\
 && \O^{AB} \cD_B \s \cD_A a_0 +\hf (\hat { \D} \s - 4) a_0 =0 ~,\\
 n a_n &+&   \O^{AB} \cD_B \s \cD_A a_{n} +\hf (\hat { \D} \s - 4) a_n = { \D} a_{n-1}~, \qquad n>0~.
 \eea
 \end{subequations}
 The bi-scalars $\s$ and $a_0$ should obey certain boundary conditions in order 
 ensure the initial condition \eqref{InitialCondition1}, including the following:
\begin{subequations}
 \bea
 \s(z,z)&=&0 ~, \quad \cD_A \s(z,z')\big|_{z=z'}=0 ; \\
a_0(z,z) &=&0~, \quad \cD_A a_0(z,z')\big|_{z=z'}=0 ~, \quad 
 \cD^2 \bar \cD^2 a_0  \s(z,z')\big|_{z=z'}= 16~. 
 \eea
 \end{subequations}
 Here we do not give the complete list of boundary conditions.
 The important for our analysis result, which follows from the above relations, is the coincidence limit 
  \bea
 U(z,z|s) = - \frac{\ri}{ (4\p s)^2} \
 \sum_{n=2}^{\infty} a_n(z,z) (\ri s)^n ~.
 \label{4.24}
 \eea
It is the coefficient $a_2(z,z)$ which contributes to the logarithmically divergent part 
of the effective action.  In the case that $\D$ coincides with the operator \eqref{3.9},
the corresponding kernel \eqref{5.44} will be denoted ${\bm U}_{\rm v}(z,z|s)$, 
and boldface notation will also be used for the bi-scalars in the right-hand side of \eqref{5.44}.

A typical contribution to the effective action is
$\G = \frac{\ri}{2} {\rm Tr} \, \ln \D$, and this can be regularised as 
\bea
\G_\o = \hf \m^{2\o} \int_0^{\infty} \frac{ \rd s }{(\ri s)^{1-\o}} 
\int \rd^4x \rd^2 \q  \rd^2 \bar \q \, E \, U(z,z|s)~,
\eea
with $\o$ the regularisation parameter. In the limit $\o\to 0$, one obtains
\bea
\G_\o = \frac{1}{32\p^2 \o} \int 
\rd^4x \rd^2 \q  \rd^2 \bar \q \, E \, a_2(z,z) + \text{finite part} 
\eea

 
\subsection{Evaluation of the heat kernel in flat superspace}

To compute the $\X$-dependent contributions from the second term in \eqref{3.16b}, it suffices to work in flat superspace and analyse the heat kernel 
\begin{subequations} \label{5.1}
\bea
 {\bm U}_{\rm v}(z,z'|s) = \re^{\ri s {\bm \D}_{\rm v} }\delta^{4|4}(z,z') ~, 
 \qquad \delta^{4|4}(z,z')=\d^4(x-x') \d^2(\q-\q') \d^2 (\bar \q - \bar \q') 
 \eea
of the operator
\bea
{\bm \D}_{\rm v} &=& \X \pa^a\pa_a 
-\frac{\ri}{2} \Big( (D^\a\X) \pa_{\a\ad} \bar D^\ad +(\bar D^\ad \X) \pa_{\a\ad} D^\a\Big)
+\frac{1}{16} \Big( (\bar D^2 \X) D^2 + (D^2 \X) \bar D^2\Big) \non \\
&& -\frac 14 \X G^{\a\ad}\big[ D_\a , \bar D_\ad \big]~,
\eea
\end{subequations}
for some real vector superfield $G_{\a\ad}$. 

When computing the heat kernel in superspace, it is convenient to deal with  
the supersymmetric interval \cite{AV}
\bea
 \z^A = 
\left\{
\begin{array}{l}
\z^a = (x-x')^a - \ri (\q-\q') \s^a {\bar \q}'
+ \ri \q' \s^a ( {\bar \q} - {\bar \q}') ~, \\
\z^\a = (\q - \q')^\a ~, \\
{\bar \z}_\ad =({\bar \q} -{\bar \q}' )_\ad ~.
\end{array} 
\right. 
\label{super-interval}
\eea
We introduce a Fourier transform for the bosonic part of the superspace delta function,
\bea
\delta^{4|4}(z,z') = \int \frac{{\rm d}^4k}{(2 \pi)^4} \, 
{\rm e}^{{\rm i} k_a \z^a} \,
\z^2 \bar{\z}^2~, \quad \z^2 = \z^\a \z_\a~, \quad \bar \z^2 = \bar \z_\ad \bar \z^\ad~.
\eea
 Making use of this representation, for the heat kernel \eqref{5.1} at coincident points we obtain
 \bea
{\bm U}_{\rm v}(z,z|s) &=& \int \frac{{\rm d}^4k}{(2 \pi)^4} \, 
\exp \, \bigg\{ \ri s \Big(
 \X X^a X_a 
-\frac{\ri}{2} \big( D^\a\X X_{\a\ad} \bar X^\ad +\bar D^\ad \X X_{\a\ad} X^\a\big) \non \\
&&+\frac{1}{16} \big(\bar D^2 \X X^\a X_\a+ D^2 \X \bar X_\ad \bar X^\ad \big) 
 -\frac 14 \X G^{\a\ad}\big[ X_\a , \bar X_\ad \big] \Big)\bigg\} \z^2 \bar{\z}^2
 \Big|_{\zeta=0}~,~~~
\label{5.4}
\eea 
where we have denoted 
\bea
X_a = \pa_a + {\rm i} k_a~, \quad 
X_\a = D_\a  - k^a (\s_a)_{\a \ad } {\bar \z}^\ad~,
\quad 
{\bar X}^\ad = {\bar D}^\ad - k^a (\tilde{\s}_a)^{\ad \a} \z_\a~.
\eea
As usual, it is useful to rescale, $k_a \to s^{-1/2} k_a$, the integration variable 
in \eqref{5.4}. Then it follows from \eqref{5.4} that  ${\bm U}_{\rm v}(z,z|s)$
 indeed has the asymptotic form \eqref{4.24}. We denote ${\bm a}_n(z,z)$ the corresponding DeWitt coefficients. 

In computing the DeWitt coefficients ${\bm a}_n(z,z)$, 
the  generic term in
the Taylor expansion of the right-hand side of \eqref{5.4}
will involve Gaussian moments of the form
\bea
\langle k^{a_1} \ldots  k^{a_n} \rangle \equiv
\frac{1}{(4 \p^2 s)^{2}} \,\int {\rm d}^4 k \,
\re^{- {\rm i} k^2 \X } \,
 k^{a_1} \ldots  k^{a_n}~.
\label{moment}
\eea
They can  be computed by introducing a generating 
function $Z(J)$ defined by 
\bea
Z(J) &=& \frac{1}{(4 \p^2 s)^{2}} \,\int {\rm d}^4 k \,
{\rm e}^{- {\rm i} k^2 \X+ J_a k^a}~, \quad
\langle k^{a_1} \ldots  k^{a_n} \rangle =
\frac{\pa^n}{\pa J_{a_1} \ldots \pa J_{a_n} }\, Z(J)\Big|_{J=0}~.
\eea
Then for $Z(J)$ one  gets 
\bea
Z(J) = \frac{\rm i}{( 4 \p {\rm i}s)^{2}} \frac{1}{\X^2}
\exp \Big( -\frac{\ri}{4\X} J^2\Big)~.
\label{genf}
\eea

The result of calculation of the coefficient ${\bm a}_2(z,z)$ is 
\bea
{\bm a}_2(z,z) &=& \frac{1}{16} 
\frac{\nabla^2 \X \bar \nabla^2 \X
- 8 G^{\a\ad} D_\a  \X \bar D_\ad \X }{\X^2} 
+\frac 18 (D \ln \X)^2 (\bar D \ln \X)^2 - G^{a} G_{a}~,
\label{5.18}
\eea
where we have denoted 
\bea
\nabla^2 \X:= D^2 \X - 2 \frac{(D \X)^2}{\X}~.
\eea
The expression \eqref{5.18} can now be lifted to curved superspace by replacing 
$D_A \to \cD_A$. 
Recalling the expression for $\X$ in terms of $\F$ and its conjugate, 
$\X=\ri (\bar \F - \F)$, we obtain
 \bea
{\bm a}_2(z,z) &=& 
-\frac{1}{16}  
\frac{1}{(\F - \bar \F)^2} 
\Big\{ \nabla^2 \F \bar \nabla^2 \bar \F -8 \cD^\a \F G_{\a\ad} \bar \cD^\ad \bar \F\Big\}
\non \\
&& + \frac{1}{8}  \frac{1}{(\F - \bar \F)^4} \cD^\a \F \cD_\a \F 
\bar \cD_\ad \bar \F \bar \cD^\ad \bar \F - G^{a} G_{a}~,
\eea
where $\nabla^2 \F $ and $ \bar \nabla^2 \bar \F $ are defined in \eqref{2.44}.
This result can also be derived directly in curved superspace by making use of either 
(i) the recurrence relations \eqref{recurrence}; or (ii) superspace normal coordinates
\cite{McA-normal}. The latter approach has recently been advocated in  \cite{Leung2}. 

We note that the  $a_2$-coefficient  for the minimal operator \eqref{operator4.9}
was computed in \cite{BK86}. The result is $a_2(z,z) =-G^a G_a$. 
Then, the logarithmically divergent 
contribution from in the second term in  \eqref{3.16b} is determined by 
\bea
{\bm a}_2(z,z) -a_2(z,z) &=& 
-\frac{1}{16}  
\frac{1}{(\F - \bar \F)^2} 
\Big\{ \nabla^2 \F \bar \nabla^2 \bar \F -8 \cD^\a \F G_{\a\ad} \bar \cD^\ad \bar \F\Big\}
\non \\
&& + \frac{1}{8}  \frac{1}{(\F - \bar \F)^4} \cD^\a \F \cD_\a \F 
\bar \cD_\ad \bar \F \bar \cD^\ad \bar \F ~.
\label{5.21}
\eea
This contribution is exactly of the type given by \eqref{action2.3}.


\subsection{Super-Weyl anomaly}

Here we compute the super-Weyl variation of the functional 
\bea
W^{(\rm v)}=  \frac{\ri}{2} {\rm Tr} \, \ln \frac{{\bm \D}_{\rm v}}{\Box_{\rm v} }~,
\eea
which is the second term in  \eqref{3.16b}. 
We are going to work with its regularised version
\bea
W^{(\rm v)}_\o = \hf \m^{2\o} \int_0^{\infty} \frac{ \rd s }{(\ri s)^{1-\o}} 
\int \rd^4x \rd^2 \q  \rd^2 \bar \q \, E \, \Big\{ {\bm U}_{\rm v}(z,z|s)
-U_{\rm v} (z,z|s)\Big\}~,
\label{5.23}
\eea
where $U_{\rm v}(s) $ is the heat kernel associated with the operator
\eqref{operator4.9}. The super-Weyl variation of ${\bm \D}_{\rm v}$ 
is readily determined by representing this operator as
\bea
{\bm \D}_{\rm v} &=& \frac{\ri}{8} \cD^\a \F (\bar \cD^2 -4R)\cD_\a
- \frac{\ri}{8} \bar \cD_\ad \bar \F ( \cD^2 -4\bar R)\bar \cD^\ad 
+\cP_{+-} \X \cP_{-+} + \cP_{-+} \X \cP_{+-}~,
\non \\
&&
\cP_{+-} := -\frac 14 (\bar \cD^2 -4R)~, \qquad 
\cP_{-+} := -\frac 14 ( \cD^2 -4\bar R)~.
\eea
where $\F$, $\bar \F$ and $\X$ are viewed as operators.
With the understanding 
that ${\bm \D}_{\rm v} $ acts on the space of scalar superfields, 
the super-Weyl variation of ${\bm \D}_{\rm v}$ can be represented as
\bea
\d_\s {\bm \D}_{\rm v} &=& (\s +\bar \s ) {\bm \D}_{\rm v} 
+(\s-\bar \s) \big( \cP_{+-} \X \cP_{-+} -\cP_{-+} \X \cP_{+-}\big)\non \\
&& - \big( \cP_{+-} \X \cP_{-+} -\cP_{-+} \X \cP_{+-}\big) (\s-\bar \s) ~.
\label{5.25}
\eea
The super-Weyl transformation of $\Box_{\rm v} $ is obtained from 
\eqref{5.25} by replacing ${\bm \D}_{\rm v} \to {\Box}_{\rm v} $ and setting 
$\X=1$. These transformations and certain Ward identities, similar to those described 
in Appendix C of \cite{BK86}, should be used to derive the super-Weyl variation of 
\eqref{5.23}. The result is 
\bea
\d_\s W^{(\rm v)}_\o &=&- \hf \o  \int_0^{\infty} \frac{\m^{2\o} \rd s }{(\ri s)^{1-\o}} 
\int \rd^4x \rd^2 \q  \rd^2 \bar \q \, E \, (\s+\bar \s) \Big\{ {\bm U}_{\rm v}(z,z|s)
-U_{\rm v} (z,z|s)\Big\}~.~~~
\eea
In the limit $\o\to 0$ we obtain 
\bea
\d_\s W^{(\rm v)}_{\rm ren} &=& -\frac{1}{32\p^2} 
\int \rd^4x \rd^2 \q  \rd^2 \bar \q \, E \, (\s+\bar \s) 
\Big\{ {\bm a}_2(z,z) -a_2(z,z) \Big\} \non \\
&=& \frac{1}{512\p^2} 
\int \rd^4x \rd^2 \q  \rd^2 \bar \q \, E \, (\s+\bar \s) 
\bigg\{
\frac{1}{(\F - \bar \F)^2} 
\Big( \nabla^2 \F \bar \nabla^2 \bar \F -8 \cD^\a \F G_{\a\ad} \bar \cD^\ad \bar \F\Big)
\non \\
&& \qquad \qquad -2  \frac{1}{(\F - \bar \F)^4} \cD^\a \F \cD_\a \F 
\bar \cD_\ad \bar \F \bar \cD^\ad \bar \F \bigg\}~.
\label{5.27}
\eea


\section{Heat kernel calculations (II)} \label{section6}

Now we turn to computing the contribution from the third term in \eqref{3.16b}. 
For this we first need to analyse the effective action of the following model 
in curved superspace
\bea
S [\J , \bar \J ; \cV]=  \int 
\rd^4x \rd^2 \q  \rd^2 \bar \q \, 
E \, \bar \J \re^{\cV} \J ~, \qquad \bar \cD^\ad \J =0~.
\label{6.1} 
\eea
Here the dynamical variables are 
 the chiral scalar superfield $\J$ and its conjugate $\bar \J$. 
 They couple to a background scalar superfield $\cV$. At the classical level, 
 the action \eqref{6.1}
 possesses two local symmetries associated with  the background fields:
 (i) 
 it
 is invariant under gauge transformations 
 \bea
 \d_\l \cV = \l +\bar \l~, \qquad \d_\l \J  = - \l \J~, \qquad \bar \cD^\ad \l =0~;
 \label{chiralgauge}
 \eea
 (ii) 
 it
 is invariant under super-Weyl transformations 
 acting on $\J$ and $V$ as follows:
 \bea
 \d_\s \J = \s \J~, \qquad \d_\s V =0~.
 \label{6.2}
 \eea
 Both symmetries are anomalous at the quantum level. The anomalies were computed in \cite{BK86,BKS}, and our discussion here will build on the results of these publications. 
  
 Effective action $\G[\X]$ is defined by 
 \bea
 \re^{\ri \G[\X] } =\int \mathsf{D}\bar \J \mathsf{D} \J 
 \exp \Big( \ri S [\J , \bar \J ; \cV]\Big)~, \qquad \X := \re^\cV
 \eea
and  can be expressed as
\bea
\G [\X] = \frac{\ri}{2} {\rm Tr} \,\ln  H_\X 
= \frac{\ri}{4} {\rm Tr}_+ \,\ln (\cP^\X_{+-}\cP^\X_{-+} ) 
+ \frac{\ri}{4} {\rm Tr}_- \,\ln (\cP^\X_{-+}\cP^\X_{+-} ) ~.
\label{6.5}
 \eea
 The operators $H_\X$ and $\cP^\X_{+-}$, $\cP^\X_{-+}$ are defined 
 in \eqref{H_X} and \eqref{H+-}, respectively. In the right-hand side of \eqref{6.5}, 
  $ {\rm Tr}_+$ denotes the  chiral trace, 
   \bea
 {\rm Tr}_+   A = \int \rd^4x \rd^2 \q \, \cE \, A(z,z) ~, 
 \qquad A(z,z') := A\delta_+(z,z')~.
\eea
   Here $A $ is an operator acting on the space of covariantly chiral scalar superfields, 
   and $\delta_+(z,z')$ is the chiral delta-function defined by \eqref{6.9}.
   
 
 \subsection{Generalised Schwinger-DeWitt representation}   
   
 As follows from the representation \eqref{6.5}, 
  $\G[\X]$ can be expressed in terms of 
 the heat kernels  ${\bm U}_{\rm c}(z,z'|s) $ and $ {\bm U}_{\rm a}(z,z'|s) $
 of the chiral  $({\bm \D}_{\rm c} )$ and antichiral $({\bm \D}_{\rm a} )$
 operators, respectively, which are  defined as 
\begin{subequations}\label{6.6}
 \bea
 {\bm \D}_{\rm c}& := &\cP^\X_{+-}\cP^\X_{-+} 
 = \frac{1}{16}  (\bar \cD^2 -4R) \X  ( \cD^2 -4\bar R) \X~, \\
 {\bm \D}_{\rm a}& :=&\cP^\X_{-+}\cP^\X_{+-} 
 = \frac{1}{16}  ( \cD^2 -4 \bar R) \X  ( \bar \cD^2 -4 R) \X~.
 \eea
 \end{subequations}
The action of $ {\bm \D}_{\rm c}$
 on a covariantly chiral scalar $\eta$, $\bar \cD^\ad \eta=0$,  is given by 
 \bea
  {\bm \D}_{\rm c} \eta &=& \bigg( \X^2 \cD^a \cD_a 
  - \frac{\ri}{4} (\bar \cD_\ad \X^2) \big\{ \cD^{\a\ad}, \cD_\a \big \}
  +\frac{1}{16} \big[ (\bar \cD^2 +4R) \X^2 \big] \cD^2 \non \\
  && -\frac{\ri}{2} \big[ \X^2 G^{\a\ad} + \hf (\bar \cD^\ad \cD^\a \X^2 )\big] \cD_{\a\ad} 
  +\frac 14 \big[ \X^2 (\cD^\a R) + \frac 14 (\bar \cD^2 \cD^\a \X^2 )\big] \cD_\a \non \\
  && + \frac{1}{16} (\bar \cD^2 -4R) \big[ \X (\cD^2 -4\bar R) \X\big]
  \bigg)\eta~, \qquad \bar \cD^\ad  {\bm \D}_{\rm c} \eta =0~.
  \label{6.88}
  \eea
  The chirality of   ${\bm \D}_{\rm c} \eta$ implies the existence of a symmetric 
  bracket, $ \J_1 \star \J_2 =  \J_2 \star \J_1$,  on the space of covariantly chiral scalars:
  \bea
  \J_1 \star \J_2& :=& \X^2 \cD^a \J_1 \cD_a \J_2 
  -\frac{\ri}{4} (\bar \cD_\ad \X^2) \big( \cD^{\a\ad} \J_1 \cD_\a \J_2 
+  \cD^{\a\ad} \J_2 \cD_\a \J_1 \big) \non \\
&&  +\frac{1}{16} \big( (\bar \cD^2 + 4R ) \X^2 \big) \cD^\a \J_1 \cD_\a \J_2~.
\label{bracket}
\eea
For arbitrary chiral scalars $\J_1 $ and $\J_2$, their bracket $ \J_1 \star \J_2$ is chiral. 
This may be checked using the algebra of  covariant derivatives \eqref{algebra}.
Setting $\X=1$ in \eqref{bracket} gives the bracket introduced in \cite{BK86,BK}.

It should be remarked that both operators \eqref{3.9} and \eqref{6.88} are non-minimal. 
Instead of dealing with $ {\bm \D}_{\rm v}$, one can equivalently work with the minimal 
symmetric operator $\widetilde  {\bm \D}_{\rm v}= \X^{-\hf}  {\bm \D}_{\rm v} \X^{-\hf}$, 
which is of the type studied in \cite{BK86,BK}.\footnote{A similar approach was pursued by Osborn in the non-supersymmetric case \cite{Osborn}.} 
However, such a transformation is not possible 
for the chiral operator $ {\bm \D}_{\rm c}$, eq. \eqref{6.88}.

Let us introduce the heat kernel associated with  the chiral operator $ {\bm \D}_{\rm c}$,
\bea
 {\bm U}_{\rm c}(z,z'|s) = \re^{\ri s {\bm \D}_{\rm c} }\delta_+(z,z')~, \qquad
\d_+ (z,z') = - \frac 14 (\bar \cD^2 - 4R) \d^{4|4} (z,z')~.
\label{6.9}
\eea
The heat kernel is chiral in both superspace  argument $z$ and $z'$. By construction, 
it obeys the equation
\bea
 \Big( \ri \frac{\pa}{\pa s} + { \bm \D}_{\rm c} \Big) {\bm U}_{\rm c} (z,z'|s) =0
 \label{HeatKernel_chiral}
 \eea
 and the initial condition 
 \bea
 {\bm U}_{\rm c} (z,z'|s\to 0) = \d_+ (z,z') ~.
 \eea
 We look for a solution to \eqref{HeatKernel_chiral} by making the ansatz 
 \bea
 {\bm U}_{\rm c}(z,z'|s) = - \frac{\ri}{ (4\p s)^2} \re^{\ri {\bm \s}_{\rm c} (z,z') / 2 s} 
 \sum_{n=0}^{\infty} {\bm a}^{\rm c}_n(z,z') (\ri s)^n ~.
 \label{6.12}
 \eea
 Here the symmetric bi-scalars $\bm \s$ and ${\bm a} _n$ are covariantly chiral 
and  obey the equations
 \bea
 &&  {\bm \s}_{\rm c}\star  {\bm  \s }_{\rm c}= 2{\bm \s}_{\rm c}~,\\
 && {\bm  \s}_{\rm c} \star {\bm a}^{\rm c}_0 
 +\hf (\hat {\bm \D}_{\rm c}{\bm \s}_{\rm c} - 4) {\bm a}^{\rm c}_0 =0 ~,\\
 n {\bm a}^{\rm c}_n &+&  {\bm \s}_{\rm c} \star{\bm a}^{\rm c}_{n} 
 +\hf (\hat {\bm \D}_{\rm c} {\bm \s}_{\rm c} - 4) {\bm a}^{\rm c}_n 
 = {\bm \D}_{\rm c} {\bm a}_{n-1}~, \qquad n>0~,
 \eea
 where we have introduced the differential opeartor
 \bea
\hat {\bm \D}_{\rm c} = {\bm \D}_{\rm c} -\frac{1}{16} (\bar \cD^2 -4R) \big[ \X (\cD^2 -4\bar R) \X\big]~.
\eea
 The bi-scalars ${\bm \s}_{\rm c}$ and ${\bm a}^{\rm c}_0$ should obey certain boundary conditions in order 
 ensure the initial condition \eqref{InitialCondition1}, including the following:
\begin{subequations}
 \bea
 {\bm \s}_{\rm c}(z,z)&=&0 ~, \quad \cD_A {\bm \s}_{\rm c}(z,z')\big|_{z=z'}=0 ; \\
{\bm a}^{\rm c}_0(z,z) &=&0~, \quad \cD_A {\bm a}^{\rm c}_0(z,z')\big|_{z=z'}=0 ~, \quad 
 \cD^2  {\bm a}^{\rm c}_0  \s(z,z')\big|_{z=z'}= -4~. 
 \eea
 \end{subequations}
The complete set of boundary condition in the $\X=1$ case is given in \cite{BK,BK86}.

The heat kernel at coincident points is 
\bea
 {\bm U}_{\rm c}(z,z|s) = - \frac{\ri}{ (4\p s)^2} 
 \sum_{n=1}^{\infty} {\bm a}^{\rm c}_n(z,z) (\ri s)^n ~.
 \label{6.18}
 \eea
It is $ {\bm U}_{\rm c}(z,z|s) $ and its antichiral twin $ {\bm U}_{\rm a}(z,z|s) $
which determine the regularised effective action 
 \bea
\G_\o [\X] &=&  \frac 14 \m^{2\o} \int_0^{\infty} \frac{ \rd s }{(\ri s)^{1-\o}} \Big\{ 
{\rm Tr}_+  {\bm U}_{\rm c}(s) + {\rm Tr}_-  {\bm U}_{\rm a}(s) \Big\} \non \\
&=& \frac 14 \m^{2\o} \int_0^{\infty} \frac{ \rd s }{(\ri s)^{1-\o}} 
\int 
\rd^4x \rd^2 \q \, \cE \, {\bm  U}_{\rm c}(z,z|s)+ \text{antichiral}~.
\eea
In order to derive the asymptotic expansion \eqref{6.18}, one does not need to assume 
that ${\bm U}_{\rm c}(z,z'|s) $ has the form \eqref{6.12}. It suffices to start 
from the definition \eqref{6.9} and then make use of the superspace 
normal coordinates \cite{McA-normal} in order to carry out calculations
similar to those employed in \cite{McA,Butter2009, Leung1,Leung2}.

The DeWitt coefficient ${\bm a}^{\rm c}_2(z,z)$ 
was computed in \cite{BKS}. It  is given by 
\begin{subequations}
\bea
{\bm a}_2^{\rm c} (z,z) &=& \frac 14  \cW^\a \cW_\a +\frac{1}{12} W^{\a\b\g}W_{\a\b\g} 
+ \frac{1}{48}  (\bar \cD^2 -4R) (G^a G_a +2 R \bar R) \non \\
&&- \frac{1}{96}  (\bar \cD^2 -4R)\cD^2 R~, 
\label{6.20a}
\eea
where 
\bea
\cW_\a = - \frac{1}{4}  (\bar \cD^2 -4R) \cD_\a \cV 
= -  \frac{1}{4}  (\bar \cD^2 -4R) \cD_\a \ln \,\X~.
\eea
\end{subequations}
Setting $\cW_\a =0$ in \eqref{6.20a} gives the DeWitt coefficient $a^{\rm c}_2(z,z)$ 
corresponding to the chiral d'Alembertian 
\bea
 {\Box}_{\rm c} 
 &=& \frac{1}{16}  (\bar \cD^2 -4R)   ( \cD^2 -4\bar R)  \non \\
 \Box_{\rm c} \eta &=& \Big\{ \cD^a \cD_a +\frac 14 R \cD^2 +\ri G^a \cD_a +\frac 14 (\cD^\a R) \cD_\a  -\frac 14  \big[(\bar \cD^2 -4R)\bar R\big]  \Big\} \eta ~,
 \eea
 with $\eta$ being a chiral scalar superfield. The coefficient $a^{\rm c}_2(z,z)$ 
 was computed for the first time in \cite{McA} and then re-derived in \cite{BK86} 
 using  an alternative technique. 

Now, the logarithmically divergent 
contribution from in the third term in  \eqref{3.16b} is determined by 
the chiral operator 
\bea
{\bm a}_2^{\rm c} (z,z) - a_2^{\rm c} (z,z) 
= \frac 14  \cW^\a \cW_\a 
\label{6.23}
\eea
and its conjugate. One may check that 
\bea
\int \rd^4x \rd^2 \q \, \cE \, \cW^\a \cW_\a
&=&\frac 14 \int \rd^4x \rd^2 \q  \rd^2 \bar \q \, E \,
\cD^\a\ln \X \cD_\a \ln \X \bar \cD_\ad \ln \X \bar \cD^\ad \ln \X  \non \\
&=& \frac 14\int 
\rd^4x \rd^2 \q  \rd^2 \bar \q \, 
E \, \frac{1}{(\F - \bar \F)^4} \cD^\a \F \cD_\a \F 
\bar \cD_\ad \bar \F \bar \cD^\ad \bar \F~.
\label{6.24}
\eea


\subsection{Chiral and super-Weyl anomalies}

Here we briefly re-derive the chiral and super-Weyl anomalies in the model \eqref{6.1}
following \cite{BK86,BKS}.

Under the chiral transformation \eqref{chiralgauge}, the operator 
 $\cP^\X_{+-}$ defined by \eqref{H+-} varies by the rule 
 \bea
 \d_\l \cP^\X_{+-}= \cP^\X_{+-}(\l +\bar \l) =  \l \cP^\X_{+-} + \cP^\X_{+-} \bar \l~.
 \eea
  Then making use of \eqref{6.6} and \eqref{6.9} leads to
 \bea
 \d_\l {\rm Tr}_+  {\bm U}_{\rm c}(s) 
 = 2s \frac{\pa}{\pa s} {\rm Tr}_+ \big( \l \, { \bm U}_{\rm c}(s) \big)
 +2\ri s {\rm Tr}_-  \big( \bar \l  \cP^\X_{-+}  {\bm U}_{\rm c}(s) \cP^\X_{+-}\big)~.
 \eea
 Due to the identity $ {\bm U}_{\rm c}(s) \cP^\X_{+-} =  \cP^\X_{+-} {\bm U}_{\rm a}(s) $, 
 we obtain
 \bea
 \d_\l {\rm Tr}_+  {\bm U}_{\rm c}(s) 
 = 2s \frac{\pa}{\pa s} \Big\{ 
 {\rm Tr}_+ \big( \l \, { \bm U}_{\rm c}(s) \big) 
 +  {\rm Tr}_- \big(\bar  \l \, { \bm U}_{\rm a}(s) \big) \Big\} ~.
 \eea
 From here we can read off the chiral anomaly
\bea
\d_\l \G_{\rm ren} [\X] = -\frac{1}{16\p^2} \int \rd^4x \rd^2 \q \, \cE \, \l \,{\bm a}_2^{\rm c} (z,z) 
+{\rm c.c.}~,
\label{6.28}
\eea
where $\G_{\rm ren} [\X] $ stands for the renormalised effective action. 
In the flat superspace limit \eqref{6.28}  reduces to the results obtained by 
Clark, Piguet and Sibold \cite{CPS} more than forty years ago.

Analogous calculations can be used to compute the super-Weyl anomaly.   
An infinitesimal super-Weyl transformation acts on $\cP^\X_{+-}$ as 
   \bea
 \d_\s \cP^\X_{+-}
 =  2\s \cP^\X_{+-} - \cP^\X_{+-} \bar \s= \cP^\X_{+-}(2\s -\bar \s) 
 ~.
 \eea
This leads to 
\bea
 \d_\s {\rm Tr}_+  {\bm U}_{\rm c}(s) 
 = s \frac{\pa}{\pa s} \Big\{ 
 {\rm Tr}_+ \big( \s \, { \bm U}_{\rm c}(s) \big) 
 +  {\rm Tr}_- \big(\bar  \s \, { \bm U}_{\rm a}(s) \big) \Big\} ~.
 \eea
As a consequence, the super-Weyl variation of the renormalised effective action is
\bea
\d_\s \G_{\rm ren}[\X] = -\frac{1}{32\p^2} \int \rd^4x \rd^2 \q \, \cE \, \s {\bm a}_2^{\rm c} (z,z)  +{\rm c.c.}
\label{6.31}
\eea
Both anomalies \eqref{6.28} and \eqref{6.31} are determined by the chiral coefficient 
${\bm a}_2^{\rm c} (z,z)$.
  
The above results allow us to compute the super-Weyl anomaly corresponding  to 
\bea
W^{(\rm c)} = -\frac{\ri}{2}  {\rm Tr} \, \ln \frac{{\bm H}_\X}{H^{(0)}  }~,
\eea
which is the third term in the effective action  \eqref{3.16b}. 
It follows from \eqref{6.31} that 
\bea
\d_\s W^{(\rm c)}_{\rm ren} = \frac{1}{32\p^2} \int \rd^4x \rd^2 \q \, \cE \, \s 
\Big\{ {\bm a}_2^{\rm c} (z,z)  - a_2^{\rm c} (z,z)\Big\} +{\rm c.c.}
\eea
Making use of the relations \eqref{6.23} and \eqref{6.24} gives
\bea
\d_\s W^{(\rm c)}_{\rm ren} = \frac{1}{512\p^2} \int \rd^4x \rd^2 \q \, \cE \, (\s +\bar \s) 
\frac{1}{(\F - \bar \F)^4} \cD^\a \F \cD_\a \F 
\bar \cD_\ad \bar \F \bar \cD^\ad \bar \F~.
\label{6.34}
\eea
  
 
 \section{Concluding comments} 
 
We are finally prepared to read off the $\F$-dependent sector of the logarithmically divergent part of the effective action \eqref{3.16b}. It is given by 
 \bea
\Big(\G_{\rm v} [\F, \bar \F] - \G_{\rm v} \Big)_{\rm div} = 
 \frac{1}{32\p^2 \o} \bigg\{ 
 \int 
\rd^4x \rd^2 \q  \rd^2 \bar \q \, E \, \Big( {\bm a}_2 (z,z) &-& a_2(z,z)\Big) \non \\
 -   \int 
\rd^4x \rd^2 \q  \rd^2  \, \cE \, \Big({\bm a}_2^{\rm c} (z,z) &-& a_2^{\rm c} (z,z) 
\Big)\bigg\}~.
\eea
Making use of the relations \eqref{5.21}, \eqref{6.23} and \eqref{6.24}, 
for the right-hand side we obtain 
\bea
- \frac{1}{512\p^2 \o} 
 \int 
\rd^4x \rd^2 \q  \rd^2 \bar \q \, E \,
\bigg\{
\frac{
\nabla^2 \F \bar \nabla^2 \bar \F -8 \cD^\a \F G_{\a\ad} \bar \cD^\ad \bar \F
}{(\F - \bar \F)^2} 
-  \frac{ \cD^\a \F \cD_\a \F 
\bar \cD_\ad \bar \F \bar \cD^\ad \bar \F 
}{(\F - \bar \F)^4} 
\bigg\}~.~~
 \eea
 This induced action is of the form \eqref{action2.3}.
 
 We are also in a position to determine the super-Weyl variation of the renormalised 
 effective action \eqref{3.16b}. Making use of the relations
 \eqref{5.27} and  \eqref{6.34} gives
 \bea
\d_\s \Big(\G_{\rm v} [\F, \bar \F] &-& \G_{\rm v} \Big)_{\rm ren}
= \frac{1}{512\p^2} 
\int \rd^4x \rd^2 \q  \rd^2 \bar \q \, E \, (\s+\bar \s)  \non \\
& \times &\bigg\{
\frac{
\nabla^2 \F \bar \nabla^2 \bar \F -8 \cD^\a \F G_{\a\ad} \bar \cD^\ad \bar \F
}{(\F - \bar \F)^2} 
-  \frac{ \cD^\a \F \cD_\a \F 
\bar \cD_\ad \bar \F \bar \cD^\ad \bar \F 
}{(\F - \bar \F)^4} 
\bigg\}~.
\eea
This anomaly is of the form \eqref{N=1SWA5}. 

The results of the last three sections can be extended to the case of $n$ vector multiplets, 
with the duality group $\sSL(2,{\mathbb R}) \cong \mathsf{Sp}(2, {\mathbb R})$ being replaced with $\mathsf{Sp}(2n, {\mathbb R})$. For certain Hermitian symmetric spaces including $\mathsf{Sp}(2, {\mathbb R})/ \mathsf{U}(1)$ and ${\mathbb C}P^n$, the curvature tensor is such  that 
\bea
R_{I \bar J K  \bar L}   \cD^\a \F^I \cD_\a \F^K \bar \cD_\ad 
\bar \F^{\bar J} \bar \cD^\ad \bar \F^{\bar L} \propto
g_{I \bar J} g_{ K \bar L}   \cD^\a \F^I \cD_\a \F^K \bar \cD_\ad 
\bar \F^{\bar J} \bar \cD^\ad \bar \F^{\bar L}~.
\eea
For a generic K\"ahler manifold $\cM$, however,
the last term in \eqref{2.4} and the following functional 
\bea
\int 
\rd^4x \rd^2 \q  \rd^2 \bar \q \, E\,g_{I \bar J} (\F, \bar \F)
g_{K  \bar L} (\F, \bar \F)  \cD^\a \F^I \cD_\a \F^K \bar \cD_\ad 
\bar \F^{\bar J} \bar \cD^\ad \bar \F^{\bar L}
\label{7.5}
\eea
are independent super-Weyl invariants. In general, \eqref{7.5} should be added to \eqref{2.4}, 
although it is not present in the $\cN=2$ case, as follows from \eqref{2.14}.
Similarly, the super-Weyl anomaly \eqref{N=1SWA5} may include, in general,
an additional contribution\footnote{The super-Weyl invariant \eqref{7.5} and anomaly contribution \eqref{7.6} were missed in the first and second arXiv versions of this paper.
I thank Adam Schwimmer and Stefan Theisen for
bringing the structures \eqref{7.5} and \eqref{7.6} to my attention.}
 \bea
 \int 
\rd^4x \rd^2 \q  \rd^2 \bar \q \, E\, (\s +\bar \s) g_{I \bar J} (\F, \bar \F)
g_{K  \bar L} (\F, \bar \F)  \cD^\a \F^I \cD_\a \F^K \bar \cD_\ad 
\bar \F^{\bar J} \bar \cD^\ad \bar \F^{\bar L}~.
\label{7.6}
\eea
 
 It would be interesting to extend the analysis of the last three sections 
  to the case of local $\cN=2$ supersymmetry. 
 Then the action \eqref{VM-action} must be replaced with 
 \bea
S [{\mathbb V}; X , \bar X]= - \frac{\ri }{8} \int \rd^4x \rd^4 \q  \, \cE \, X 
\Big(W ({\mathbb V}) \Big)^2
+{\rm c.c.}
\label{VM-actionN=2}
\eea
Here $X $ is a background chiral scalar superfield containing the dilaton and axion as the lowest component, and $W$ is the field strength of a vector multiplet. The latter is a reduced chiral superfield, 
 \bea
{\bar \cD}^{\ad }_i W =0~, \qquad 
\Big(\cD^{ij}+4S^{ij}\Big)W
=
\Big(\cDB^{ij}+ 4\bar{S}^{ij}\Big)\bar{W}~,
\label{vectromul}
\eea
see \cite{KLRT-M} for the technical details. There are three different realisations for the unconstrained prepotential $\mathbb V$ in \eqref{VM-actionN=2}.
 One option  is to introduce a curved-superspace extension 
of Mezincescu's prepotential \cite{Mezincescu} (see also \cite{HST}),  $V_{ij}=V_{ji}$,
which is an unconstrained real SU(2) triplet. The expression for $W$ in terms of $V_{ij}$ 
was derived in \cite{ButterK11}, and is given by 
\bea
W = \frac{1}{4}\bar\Delta \Big({\cD}^{ij} + 4 S^{ij}\Big) V_{ij}~,
\eea
where $\bar\Delta$ is the chiral projection operator  \cite{Muller,KT-M09}.
Another option is the analytic prepotential $V^{++} $ which originates within the harmonic superspace approach \cite{GIKOS,GIOS}. Finally, one can work with the tropical prepotential $V(\z)$ corresponding to the projective superspace approach 
\cite{KLR,LR1,LR2}. It is not completely obvious which of the three prepotential is the best choice to perform loop calculations in supergravity.
\\

 
\noindent
{\bf Acknowledgements:}\\
I am grateful to 
Stefan Theisen and Arkady Tseytlin for discussions, and to Darren Grasso and Ian McArthur for comments on the manuscript. 
This work  is supported in part by the Australian Research Council, project No. DP200101944.

 
 \appendix 
 
 \section{Super-Weyl transformations}

The simplest approach to describe $\cN=1$ conformal supergravity in superspace 
is to make use 
of the Grimm-Wess-Zumino geometry \cite{GWZ}, which is at the heart of
the Wess-Zumino formulation for old minimal supergravity \cite{WZ}, 
in conjunction with the super-Weyl transformations  \cite{Siegel78,HT}.
 The geometry of curved superspace is described by covariant derivatives
of the form
\bea
\cD_{A}=(\cD_{{a}}, \cD_{{\a}},\cDB^\ad)
=E_{A}{}^M \pa_M
+\hf \O_{{A}}{}^{bc}M_{bc} 
~,
\label{CovDev}
\eea
which obey the graded commutation relations \cite{BK}
\begin{subequations}\label{algebra}
\bea
& \{ \cD_\a , {\bar \cD}_\ad \} = -2{\rm i} \cD_{\a \ad} ~,\\
&
\{\cD_\a, \cD_\b \} = -4{\bar R} M_{\a \b}~,
 \qquad
\{ {\bar \cD}_\ad, {\bar \cD}_\bd \} =  4R {\bar M}_{\ad \bd}~, 
 \\
&\left[ \cD_{\a} , \cD_{ \b \bd } \right]
      = 
     {\rm i}  {\ve}_{\a \b}
\Big({\bar R}\,\cDB_\bd + G^\g{}_\bd \cD_\g
- \cD^\g G^\d{}_\bd  M_{\g \d}
+2{\bar W}_\bd{}^{\gd \dot{\d}}
{\bar M}_{\gd \dot{\d} }  \Big)
+ {\rm i} \cDB_\bd {\bar R} \, M_{\a \b}~,
~~~~~~\\
&\left[ { \bar \cD}_{\ad} , \cD_{ \b \bd } \right]
      =  -{\rm i}{\ve}_{\ad \bd}
\Big(R\,\cD_\b + G_\b{}^{\dot{\g}}  \cDB_{\dot{\g}}
-\cDB^\gd G_\b{}^{\dot{\d}}
{\bar M}_{\gd \dot{\d}}
+2W_\b{}^{\g \d}
M_{\g \d} \Big)
- {\rm i} \cD_\b R  {\bar M}_{\ad \bd}~.~~~~~~~ ~~
\eea
\esubeq
Here the torsion tensors $R$, $G_a = {\bar G}_a$ and
$W_{\a \b \g} = W_{(\a \b\g)}$ satisfy the  Bianchi identities:
\begin{subequations}
\bea
&\cDB_\ad R= 0~,~~~~~~\cDB_\ad W_{\a \b \g} = 0~,
\\
&
\cDB^\gd G_{\a \gd} = \cD_\a R~,~~~~~~
\cD^\g W_{\a \b \g} = {\rm i} \,\cD_{(\a }{}^\gd G_{\b) \gd}~.
\eea
\end{subequations}

The infinitesimal super-Weyl transformation is given by 
\begin{subequations} 
\label{superweyl}
\bea
\d_\s \cD_\a &=& ( {\bar \s} - \hf \s)  \cD_\a + \cD^\b \s \, M_{\a \b}  ~, \\
\d_\s \bar \cD_\ad & = & (  \s -  \hf {\bar \s})
\bar \cD_\ad +  ( \bar \cD^\bd  {\bar \s} )  {\bar M}_{\ad \bd} ~,\\
\d_\s \cD_{\a\ad} &=& \hf( \s +\bar \s) \cD_{\a\ad} 
+\frac{\ri}{2} \bar \cD_\ad \bar \s \,\cD_\a + \frac{\ri}{2}  \cD_\a  \s\, \bar \cD_\ad
+ \cD^\b{}_\ad \s\, M_{\a\b} + \cD_\a{}^\bd \bar \s\, \bar M_{\ad \bd}~.
~~~~~~
\eea
\end{subequations}
It generates 
the following transformations of the torsion superfields
\begin{subequations} 
\bea
\d_\s R &=& 2\s R +\frac{1}{4} (\bar \cD^2 -4R ) \bar \s ~, \\
\d_\s G_{\a\ad} &=& \hf (\s +\bar \s) G_{\a\ad} +\ri \cD_{\a\ad} ( \s- \bar \s) ~, 
\label{s-WeylG}\\
\d_\s W_{\a\b\g} &=&\frac{3}{2} \s W_{\a\b\g}~.
\label{s-WeylW}
\eea
\end{subequations} 
Here the super-Weyl parameter $\s$ is a covariantly chiral scalar superfield,  $\bar \cD_\ad \s =0$. The super-Weyl transformations belong to the  
gauge group of conformal supergravity.

\begin{footnotesize}

\end{footnotesize}


\end{document}